# Plenty of motion at the bottom:
# Atomically thin liquid gold membrane

**Authors:** Pekka Koskinen[1]* and Topi Korhonen[1]

**Affiliations:**

[1]Nanoscience Center, Department of Physics, University of Jyvaskyla, 40014 Jyväskylä, Finland.

*Correspondence to: pekka.koskinen@iki.fi

**Abstract:** The discovery of graphene some ten years ago was the first proof of a free-standing two-dimensional (2D) solid phase. Here, using quantum molecular dynamics simulations of nanoscale gold patches suspended in graphene pores, we predict the existence of an atomically thin, free-standing 2D liquid phase. The liquid phase, enabled by the exceptional planar stability of gold due to relativistic effects, demonstrates extreme fluxionality of metal nanostructures and opens possibilities for a variety of nanoscale phenomena.

Solid and liquid are familiar phases, but they usually refer to three-dimensional materials. The discovery of graphene proved that materials can exist also in a two-dimensional (2D) solid phase, which was until then considered unfeasible (*1*, *2*). Later graphene has been followed by other 2D materials like hexagonal boron nitride and transition metal dichalcogenides, along with a plethora of new physics and applications (*3–10*). These 2D materials are characterized by strong covalent bonding within layers and by weak van der Waals bonding between successive layers. The covalent bonds are directional, which means that atoms have rigid positions and they move only when subject to high temperatures or to irradiation by electrons or ions (*11*).This 2D directional rigidity implies that 2D covalent materials cannot exist in a liquid phase.

However, recent experiments demonstrated that the 2D solid phase could exist also in metals, at least in a suspended nanocrystal. The demonstration was done by creating atomically thin and free-standing metal patches suspended in graphene pores (*12*). The metal in the experiment was iron, but gold would be another particularly suitable patching metal for two reasons. First, the interaction of gold with graphene is suitable. Gold (Au) diffuses swiftly on top of graphene (*13*), decorates readily bare graphene edges (*14*, *15*), and shows strong in-plane binding to graphene due to interaction between the d-orbitals of Au and the **π**-electron cloud of graphene (*16*). Second, and more important, Au among all metals shows an exceptional propensity for planar structures, which could enable relatively large stable 2D patches (*17–21*). Here we investigate such Au patches in graphene pores by quantum molecular dynamics simulations. It turns out that, compared to the covalent 2D materials with rigid structures, the flexible 2D metallic bonding facilitates atomic motion of quite different nature.

To create a model for a suspended 2D Au membrane, we embedded an Au$_{49}$ patch inside a graphene pore of a suitable size (Fig. 1A). This patch was roughly of the same size as the experimental Fe patches (*12*). After constructing the system, it was optimized and simulated by quantum molecular dynamics (MD) simulations at select temperatures. The electronic structure was calculated by density-functional tight-binding (DFTB), which reproduces the bonding trends of Au with sufficient accuracy for our qualitative purposes (*22*). For discussion about methods, simulation details, and convergence analysis, see Supplementary Information. Structural optimization with all sensible initial guesses resulted in a 2D close-packed nanocrystal with mild distortions due to interaction with graphene edges. The optimization was followed by MD simulations at different temperatures. At 300 K the patch remained stable, with both C and Au vibrating around their equilibrium positions without diffusion (Fig. 1B). At 500 K and 700 K the vibrations intensified, but the solid phase remained. However, at 900 K Au atoms started diffusing within the plane (Fig. 1C). Au atoms bound covalently to C stood still, but atoms in the middle of the patch diffused around, swapped places, vibrated in and out of the plane but did not leave the plane. This behavior can be identified as a 2D liquid phase, and is best witnessed by the Supplementary Movie 1.

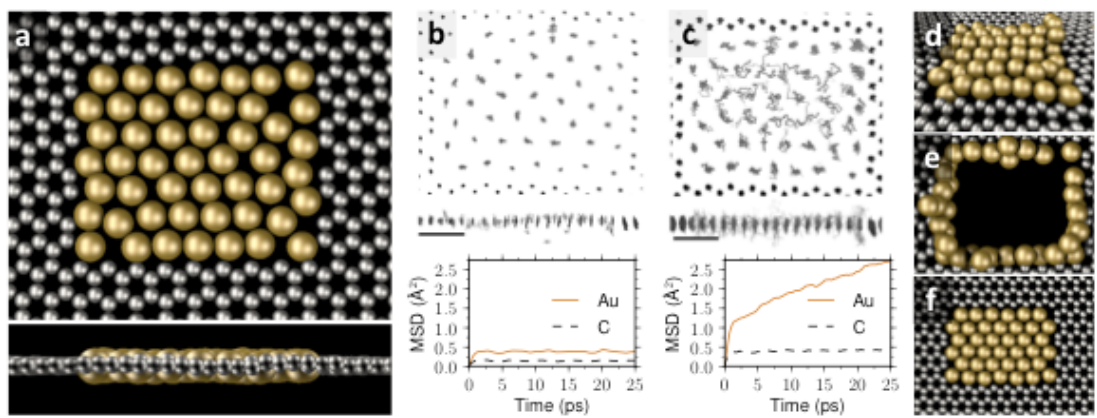

**Fig. 1**. Patching graphene pore with gold. (**A**) Optimized $Au_{49}$ patch in a graphene pore of 136 removed C atoms (top and side view). (**B**) Dynamics of $Au_{49}$ at 300 K: top and side views of Au and C atom trajectories (10 ps duration), along with mean square displacements (MSD) calculated from 0.1 ns simulation. Patch is in solid phase with no diffusion. (**C**) The same as panel **B** at 900 K. The patch is in liquid phase and the MSD slope of Au corresponds to diffusion constant 0.016 $Å^2$/ps. (**D**) Snapshot of $Au_{49}$ at 900 K. As a detail, see the Au atoms piled at the upper right hand corner. (**E**) Snapshot of a ruptured $Au_{48}$ membrane. (**F**) Optimized $Au_{39}$ in a smaller graphene pore of 119 removed C atoms.

While the $Au_{49}$ patch showed a stable liquid phase, some other patches did not. We repeated the simulation by filling the same graphene pore by $Au_{40}$, $Au_{47}$ and $Au_{48}$ patches, all of which showed stable solid phases, but ruptured before melting upon temperature increase (Fig. 1E). The reason was that with fewer atoms also Au patches began to show pores that fast grew bigger and did not get self-repaired. In contrast, $Au_{49}$ patch had so many Au atoms that some of them piled at the edges (Fig. 1D) and acted as a feedstock that could—if needed—provide Au atoms to fill and self-repair small pores in the patch. This surplus of Au atoms thus made the patch and the liquid phase more stable. In addition, the pore also needed to be large enough. In fact, the above simulations were preceded by simulations of a smaller pore patched by $Au_{39}$ that also showed a stable solid phase (Fig. 1F). However, now the Au atoms were too tightly constrained because the pore was too small, and the patch could not show liquid phase. In this case even changing the number of Au atoms did not help. The patch simply ruptured upon heating. These results imply that in experiments the pores should be large enough and the pore should be completely filled by atoms before increasing the temperature. However, these computer simulations with their limited sample sizes and especially insufficiently short time scales can say only so much about the practical stability of liquid patches. For that, experiments would be required.

In general, the existence of 2D liquid phase requires three conditions. First, the pore template itself has to remain stable at high temperatures, a condition easily met by graphene (*23*). Second, edge interactions need to favor planar bonding and be robust enough to endure high temperatures. Our supplementary calculations showed that the Au-C interface has bending rigidity comparable to that of the 2D Au membrane, which is sufficient to retain the patch steady under Au diffusion (Fig. S4). Third, the membrane itself has to display 2D diffusion before out-of-plane fluctuations grow too large and initiate rupturing. The first two requirements are related to the experimental setting of a patched pore. They could be investigated more systematically for different pore and patch sizes, different template materials, and also different patching metals. In what follows, however, we wish to focus on the last condition and investigate the intrinsic properties of the 2D Au liquid, without the details of the actual experimental settings.

To model the 2D Au membrane, we performed unconstrained simulations in a cell periodic in lateral directions and non-periodic in vertical direction. We used cells small enough to represent patches in graphene pores but large enough to avoid artifacts due to periodicity; note that the underlying experimental context makes the thermodynamic limit irrelevant. Therefore most simulations were done in a 64-atom cell, which was of similar size as the patches in graphene pores and also converged with respect to diffusion behavior and the local nature of the 2D liquid phase (see Supplementary Information for convergence analysis). The two governing parameters of the system were temperature and atom density. The density was set by area strain $\alpha = (A - A_0)/A_0$, where $A$ is the lateral area of the cell and $A_0$ its equilibrium value at zero temperature. Changing $\alpha$ enabled us to imitate membranes under varying tensile strain, which could be created by filling different pores by different number of Au atoms. We conducted a systematic set of 182 simulations at $500 - 1800$ K temperatures and $0 - 12$ % area strains; each simulation consisted of equilibration and 0.25 ns of data collection (see Supplementary Information).

The simulations showed four phases (Fig. 2A). First, low temperature and small area strain resulted in a 2D solid phase, with atoms vibrating in- and out-of plane about their equilibrium positions (Fig. 2B). The bending modulus of this phase was 0.45 eV, almost half of the modulus of graphene, and it increased upon heating (see Supplementary Information). Second, low temperatures but area strains above ~ 6 % resulted in a solid phase with pores in the otherwise close-packed structure (Fig. 2C). Third, at 900-1200 K the solid phase melted to the 2D liquid phase. In this phase atoms diffused in-plane and vibrated out-of-plane, sometimes even making short-lived hops out of the plain, while still retaining the planar structure (see trajectory side view in Fig. 2D). Increasing area strain above zero decreased the melting temperature, because the tensile strain created more freedom for atoms to diffuse and move about. However, above $\alpha \approx 4$ % the area strain had only little effect on melting temperature, because lateral freedom was presumably saturated and even though pores grew larger, the average bond lengths away from the pore remained unchanged. Supplementary calculations, motivated by sheer curiosity, showed that in small cage-like Au clusters the presence of curvature further decreased the melting temperature (Fig. S5). Under all conditions the melting temperatures remained below the 3D bulk Au value of 1337 K. Fourth, at high temperatures above 1400-1700 K the 2D liquid became unstable against out-of-plane motion, atoms began to cluster into 3D bonding configurations and the 2D liquid got ruptured. The instability temperature decreased for increased area strain, in qualitative agreement with previous simulations that showed rupture of Au patches with fewer atoms.

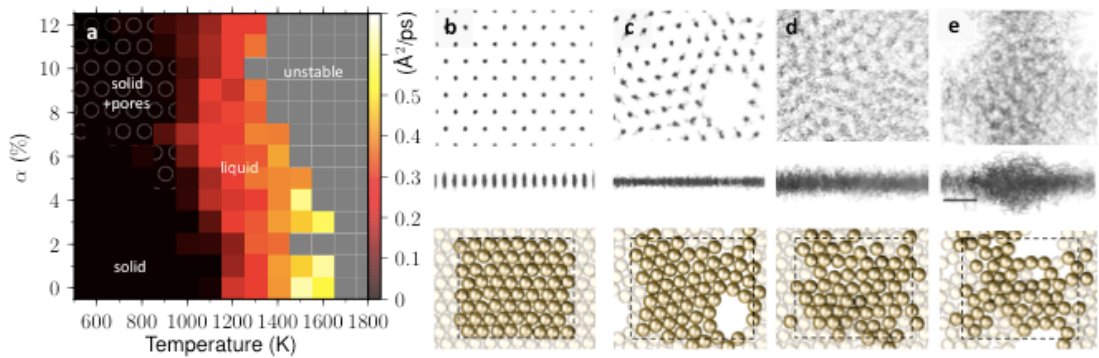

**Fig. 2**. The four phases of a 2D Au membrane. (**A**) Phase diagram as a function of temperature and area strain, plotted by using diffusion constant (colorbar). (**B**) Solid phase at 500 K and zero area strain. (**C**) Solid phase with pores at 500 K and 11 % area strain. (**D**) 2D liquid phase at 1300 K and zero area strain. (**E**) Unstable, 3D phase at 1400 K and 10 % area strain. Top and middle panels are top and side views of 50 ps atom trajectories. Bottom panels are snapshots after 25 ps equilibration: before equilibration the opaque atoms were close-packed inside the simulation cell (dashed boxes); the transparent atoms are their periodic copies. Scale bar, 0.5 nm.

The size of the simulation cell had only minor effect on the melting temperature, because the diffusion events were local. However, what the cell size did affect was the boundary of rupturing instability: 2D liquid phase was stable when the cell was sufficiently small, but it became unstable when the cell was too large. For example, at 1100 K and 5 % area strain an $Au_{256}$ system retained the 2D liquid phase with the same internal structure as $Au_{64}$. This large system further indicated that the 2D liquid phase was not an artifact of the periodic boundary conditions, but an intrinsic property of the material. However, for $Au_{400}$ system at the same temperature and area strain the out-of-plane fluctuations ultimately grew too large and the membrane ruptured. Soap bubbles are subject to similar limitations: small enough bubbles remain stable for long times, while large bubbles burst rapidly. This implies that an optimally stable 2D liquid has to balance between two competing effects: too small patches constrain planar diffusion and too large patches make the liquid phase dynamically unstable because of too large out-of-plane fluctuations. Nevertheless, the simulations suggest that, at least for experimentally relevant patch sizes, there should exist a temperature window where the liquid phase would be stable without rupturing. As the patch size increases the window narrows down and ultimately closes; an infinite free-standing 2D liquid membrane does not exist. Regarding the stability of solid patches, however, an analysis similar to that in Ref. (*12*) showed that it should be possible for Au patches to remain stable up to ~20 nm-diameter patches, much larger than for Fe (see Supplementary Information). This suggests that Au indeed could be a promising patching metal material.

A closer investigation of the intrinsic atomic dynamics of 2D Au membrane at zero area strain and different temperatures revealed typical signatures of liquid behavior (see Supplementary Information for analysis methods) (*24*). In the solid phase the velocity autocorrelation showed correlated vibrations up to few picosecond time scales (Fig. 3A). Melting made these long-time correlations disappear, and only the spatially local correlations below ~1 ps remained. At 1200 K the diffusion constant was 0.2 Å$^2$/ps, which combined with the 1 ps decorrelation time scale infers that, on average, atoms undergo couple of vibration periods before diffusing to neighboring sites. In addition to temporal dynamics, we used the pair correlation function to analyze spatial structure (Fig. 3B). In the solid phase the pair correlations extended well beyond 1 nm, but upon melting these distant correlations disappeared and the pair correlation function converged rapidly towards uncorrelated continuum limit. In the liquid phase only the nearest- and next-nearest neighbor correlations below ~7 Å remained prominent (Fig. 3C).

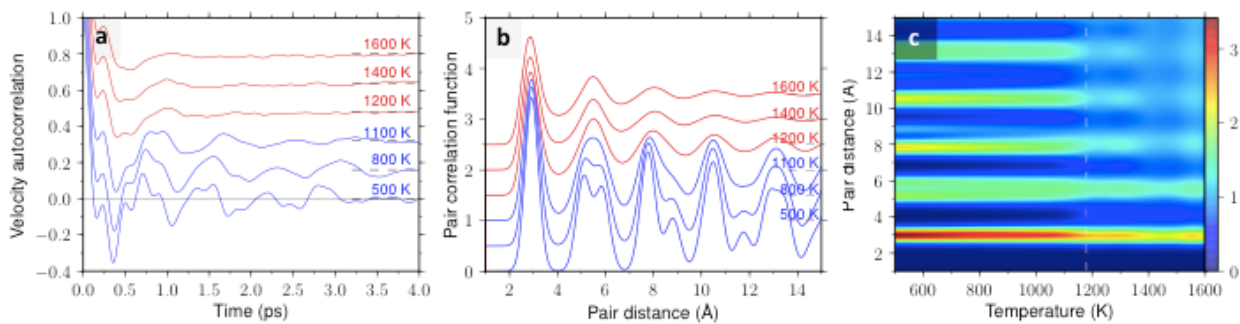

**Fig. 3.** Structure analysis of the 2D Au membrane. (**A**) Velocity autocorrelation function at different temperatures. The 500 K curve is centered at zero, others are offset for clarity. (**B**) Pair correlation function at different temperatures in solid (blue) and liquid (red) phases. The dashed lines correspond to the uncorrelated continuum limit. (**C**) Contour plot of the pair correlation function, showing the disappearance of long-range pair correlations upon melting below 1200 K. The area strain is zero in all plots.

To support the DFTB results, we simulated the same periodic Au$_{64}$ system using both DFTB and density-functional theory (DFT); see Supplementary Information for method details. Also DFT predicts the existence of the 2D liquid phase (Supplementary Movies 2 and 3). At 1600 K the diffusion constant was 0.14 Å$^2$/ps with DFT and 0.55 Å$^2$/ps with DFTB. This is a reasonable agreement, remembering that the constant depends exponentially on the diffusion energy barriers, but it suggests that the DFTB phase diagram underestimates the temperature scale. The different diffusion rates are reflected in the trajectories that show more crystallinity in DFT than in DFTB (Figs. 4 A and B). Despite these quantitative differences, the 2D liquid phase in both methods is unmistakable. Trajectory side views show that DFT shows even greater planar stability (Fig. 4, A and B). This is reasonable, as our DFT exchange-correlation functional is known to give slight overbinding of 2D gold clusters compared to 3D ones (*20*). The actual planar stability of the liquid phase probably lies somewhere in between these two results.

The exceptional planar stability of Au, at least in small and static gold clusters, is known to originate from relativity (*17*, *20*, *25*, *26*). With this in mind, we repeated the above simulations by removing relativistic effects: we used a nonrelativistic approach to solve Au atomic orbitals and energies before making DFTB parametrizations and DFT projector augmented wave setups (see Supplementary Information). Without relativity, the system of Au$_{64}$ at 1600 K and zero area strain became unstable against out-of-plane motion already within few ps, both in DFTB and DFT (Figs. 4 C and D). This instability is a strong indication that, by reducing energy gaps and increasing hybridization between 6s and 5d orbitals, relativistic effects indeed are behind the unexpected stability of the 2D liquid phase (*20*). It is good to bear this origin in mind when considering the 2D liquid phase with other metals, because relativistic effects are known to be particularly prominent in Au. Many more metals will probably make stable 2D solids, but stable 2D liquids are less likely; further investigations are required.

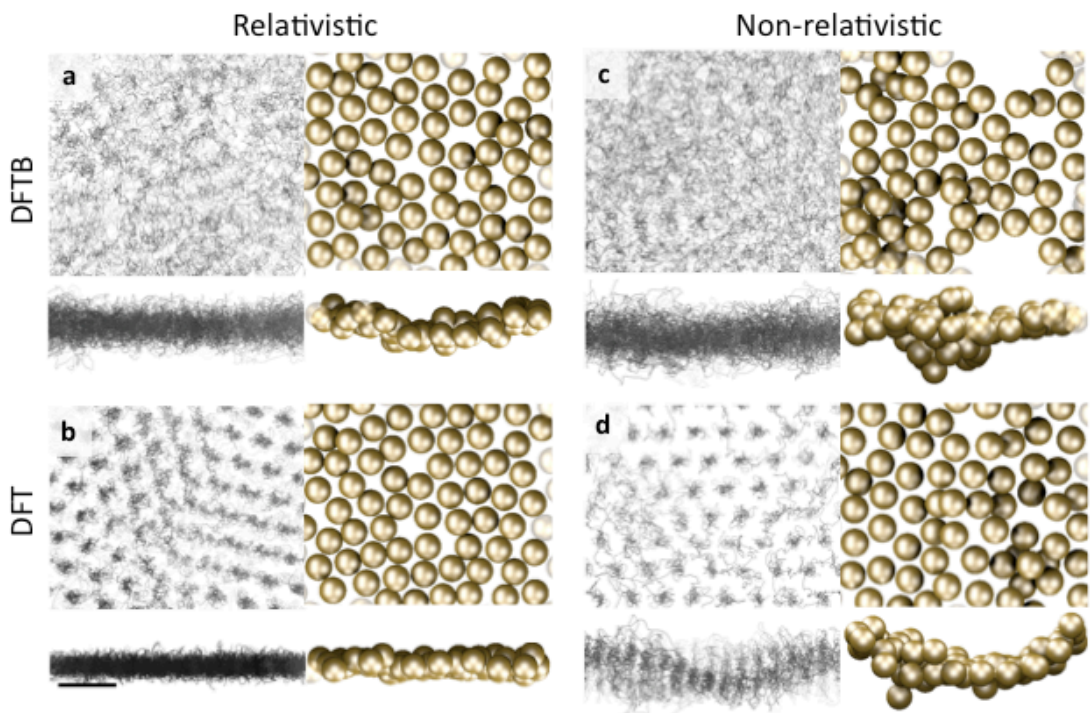

**Fig. 4**. Comparing DFT and DFTB simulations and the effect of relativity. DFTB (top panels **A** and **C**) and DFT (bottom panels **B** and **D**) of the same system and initial structure (64 atoms at 1600 K in a 23.5 Å×20.4 Å cell). Subpanels include top and side views of snapshots and atom trajectories (65 ps for relativistic, 25 ps for nonrelativistic). Scale bar, 0.5 nm.

Here we investigated 2D liquid in the context of Au-patched graphene, but the context could be expanded. Templates could be tested also with other van der Waals solids, and the patching metal could be varied. In addition to patched pores, 2D liquids might be realizable in quantum point contacts or other narrow 2D wires (*27*). Sandwiching could further stabilize the 2D liquids, although that might compromise the free-standing nature of the liquid (*28*). Also, while the temperatures appear high, the realization of liquid-like behavior should be possible also at lower temperatures under ion or electron irradiation due to beam-enhanced mobility of atoms (*12*, *16*, *29*, *30*). Alas, simulations suggest that the 2D liquid is fragile, and this certainly causes great experimental challenges.

Two-dimensional liquids have often been modeled using classical model potentials with $\sim 10^5$ atoms, focusing on theories about 2D melting (*31*, *32*), on critical or cooperative phenomena (*33*), or on other properties of phase transitions (*34*). The experimetal and theoretical contexts have been cell membranes (*35*), molten surfaces of clusters and bulk metals (*36*), small metal clusters (*37*), confined water (*38*, *39*), and even the flow of human crowds (*40*). Soap bubble is 2D liquid, too, but here it is the atomic scale that makes the distinction. Atomically thin liquid phase would provide a model system to study various material properties. Doping patches by impurity atoms would give means to track free 2D diffusion. Elongated patches or even one-dimensional metallic seams could be envisaged for the investigation of quasi-one- and two-dimensional diffusion. The malleability implied by the liquid phase could even help improving the design of mechanically durable nanostructures. Still, at nanoscale the distinction between solid and liquid phases is not clear-cut (*41*): even below melting temperature the 2D liquid phase implies structural ultrafluxionality that could provide new ideas for the design of nanostructures with customized catalytic properties (*42*).

**Acknowledgments:** We thank Gerrit Groenhof for useful comments on the manuscript. We thank the Academy of Finland for funding (projects 283103, 251216) and the CSC – IT Center for Science in Finland for computational resources.

# Supplementary Information for
# Plenty of motion at the bottom: Atomically thin, free-standing liquid gold

**Authors:** Pekka Koskinen[1]* and Topi Korhonen[1]

**Affiliations:**

[1]Nanoscience Center, Department of Physics, University of Jyvaskyla, 40014 Jyväskylä, Finland.

*Correspondence to: pekka.koskinen@iki.fi

**Materials and Methods:**

**Density-functional theory (DFT) simulations** – We used the exchange-correlation functional of Perdew, Burke, and Ernzerhof and the real-space grid DFT code GPAW (*43–45*). The code was used in the local basis mode (LCAO) with a double-zeta polarized basis, which speeded up the computations but still gave accuracy comparable to the full grid for our purposes. Grid spacing was 0.2 Å, Fermi broadening was 0.05 eV, and the k-space was sampled at the $\Gamma$-point. The vertical dimensions in the simulation cells were non-pediodic and their lengths were 16 Å. The lattice constant for infinite and close-packed 2D Au was 2.8 Å. Relativistic DFT simulations used the standard projector augmented wave (PAW) setups generated by scalar-relativistic calculation of the Au atom, as usual (*46*). Nonrelativistic DFT simulations used PAW setups and LCAO basis functions that were generated by a similar procedure with default parameters, apart from the Au atom that was solved using nonrelativistic formalism.

**Density-functional tight-binding (DFTB) simulations** – DFTB derives its parametrizations directly from *ab initio* electronic structure methods without semi-empirical fitting (*47, 48*), and is the method of choice when long-running molecular dynamics simulations are required. The DFTB Au parametrizations, established earlier, turned out to provide a fair description of the electronic structure in small nanostructures (*22*). There are quantitative differences, but the accuracy suffices for our qualitative purposes. In particular, parametrizations faithfully reproduce the propensity for planar bonding, which originates from enhanced 5d-6s orbital hybridization caused by relativistic contraction of core orbitals. This planar bonding and its origin, known from experiments and explained by DFT calculations, are reproduced also by DFTB (Fig. S1) (*22, 26*). Relativistic DFTB simulations used the parametrizations of Ref. (*22*), generated by scalar-relativistic calculation of the Au atom, just as with PAW setup generation for DFT. The relativistic DFTB parametrization gave 2.9 Å for the lattice constant of infinite close-packed 2D Au. Nonrelativistic DFTB calculations used parametrizations that were generated by the standard procedure of Ref. (*48*), apart from the Au atom that was solved using nonrelativistic formalism (repulsion was fit to $Au_2$ dimer and Au bulk with a 2.9 Å cutoff).

The DFTB C parametrizations were adopted from Ref. (*47*) and Au-C parametrizations were made following the standard procedure of Ref. (*48*): The parameters for quadratic confinements were $r_0 = 2.52$ Å for Au and $r_0 = 1.41$ Å for C, and the repulsion (with 2.65 Å cutoff) was fitted to dimer (AuC) and Au-benzene ($C_6H_5Au$) structures of varying Au-C bond lenghts (with PBE-

DFT energies). As benchmark examples, the adsorption energy of Au in graphene single vacancy was 3.4 eV in DFTB and 3.0 eV in DFT (*16*), and the adsorption energy of Au in the zigzag edge of graphene was 3.9 eV in DFTB and 3.4 eV in DFT (*15*). The agreement is satisfactory, given that also DFT numbers are functional-dependent. For systems with both Au and C atoms we used self-consistent charge (SCC) DFTB (*49*). For systems with only Au we could use a non-SCC approach without compromising accuracy because charge transfer was so small.

**Molecular dynamics (MD) simulations** – MD simulations used a 2.5 fs time step (5 fs for systems with Au only) and a Langevin thermostat. The thermostat used a 5 ps damping time (proportional to the inverse of friction coefficient), which was sufficiently large to rid all thermostat-induced artifacts in the atomic diffusion (Fig. S2A). Regarding atom diffusion in the periodic 2D Au, the most critical parameter was the lateral size of the cell. A cell with 64 atoms (size 23.5 Å×20.4 Å) showed convergent behavior with respect to both diffusion and mean cohesion (Figs. S2, B and C). All 2D Au simulations begun with 25 ps equilibration and completed with 0.25 ns data collection. Diffusion constant $D$ was calculated via mean square displacement $\langle r^2 \rangle = 4Dt$ for the selected set of atoms (*50*). The velocity autocorrelation

$$z(t) = \frac{\langle \vec{v}(t_0) \cdot \vec{v}(t - t_0) \rangle}{\langle \vec{v}(t_0) \cdot \vec{v}(t_0) \rangle},$$

where $\vec{v}(t)$ is the $3N$-dimensional velocity vector, was averaged over 50 randomized initial times $t_0$ within the given trajectory. The pair correlation function was calculated as $g(r) = \sum_{ij} \delta_d(r - r_{ij}) / (2\pi r \sigma)$, where $\sigma$ is the atom density and $\delta_d(r)$ is a Gaussian with a broadening $d = 0.25$ Å.

**Bending rigidity of 2D Au and Au-C interface** –The bending rigidity of the 2D Au was calculated by three approaches. First, Au tubes with radii of curvature below $R = 1.2$ nm were calculated both by DFT and DFTB (10 k-points for a tube of 5.1 Å periodic length). DFTB appeared to give a reasonable, even if slightly underestimated bending rigidity of 2D Au membrane when compared to DFT (Fig. S3A). This is in line with Fig. 4 in the main text. Second, following the procedure of Refs. (*51*) and (*52*), revised periodic boundary conditions (RPBC) were used to simulate the bending of an infinite 2D Au (one-atom cell with 10×10 k-points)(*53*). Fitting to radii of curvature above 20 Å yields static bending modulus of 0.45 eV, and extrapolating to radii of curvature below 20 Å yields the energies of Au tubes in satisfactory agreement (considering that tubes contain finite-size effects absent in RPBC simulations). Third, RPBC approach with $Au_{64}$ was used to calculate the temperature-dependence of the bending modulus. The modulus at given temperature was calculated by fitting the $R^{-2}$-behavior to a set of 0.5 ns simulations at $R =$ 20, 30, 50, 100, and 200 Å. Strain contributions that arose from slow radial movement of the membrane were removed prior to the fitting (the elastic modulus of 2D Au was 5.71 eV/Å$^2$). As a result, the bending rigidity increased at the rate 0.1 meV/K upon increasing temperature (Fig. S3B). Bending rigidity of the molten phase was not calculated, because it would have required inaccessibly long simulations. The bending rigidity of Au-C interface, as estimated from a periodic slab calculation, was approximately the same as that of 2D Au (Fig. S4).

**Size-dependent stability analysis for patches suspended in graphene pores** – Following Ref. 12, the stability of patches was estimated by the energy difference of pore-suspended patch and a three-dimensional nanocluster. The energy difference of a square patch of edge length *L* and a cubic nanocluster with the same number of atoms is

$$\Delta E = L^2(\varepsilon_{2D} - \varepsilon_{3D})/A_0 + 4L\varepsilon_{if} - 6(V_0/A_0)^{\frac{2}{3}}L^{\frac{4}{3}}\varepsilon_{surf}$$

where (using DFTB values) $\varepsilon_{2D} = -3.3$ eV is the 2D binding energy per atom, $\varepsilon_{3D} = -3.6$ eV is the 3D binding energy per atom (FCC bulk), $A_0 = 7.3$ Å$^2$ is the area per atom in 2D, $V_0 = 23$ Å$^3$ is the volume per atom in 3D (FCC bulk), $\varepsilon_{if} = -0.7$ eV/Å is the binding energy of Au-C interface (averaged over few randomized interfaces), and $\varepsilon_{edge} = 0.08$ eV/Å$^2$ is the surface energy (average over different facets). At small sizes the patch is energetically favorable ($\Delta E < 0$) due to the interface energy that is negative and linear in $L$. At larger sizes the 3D cluster becomes increasingly more favorable because of the 3D bulk term. The critical size above which the 3D clusters become energetically more favorable ($\Delta E = 0$) occurs around $L \approx 20$ nm.

**Curvature-dependent melting temperature** – The effect of curvature on the melting temperature of 2D Au was investigated by simulating Au cage clusters with 18, 32, 50, and 72 atoms. Unlike carbon fullerenes, the biggest of these golden cage-like isomers are most likely metastable and thus not seen in experiments (*19, 54, 55*). However, MD simulations at various temperatures showed that within 0.1 ns the clusters retained their cage-like geometries even though they were in liquid phase at the same time (Fig. S5A). Moreover, cage-like clusters showed reduced melting temperatures, apparently because in a curved surface the atoms had more freedom to move towards the convex side and thus they could overcome diffusion barriers more easily. This decrease in the melting temperature turned out to be directly proportional to the curvature (Fig. S5B).

**Figures S1-S5:**

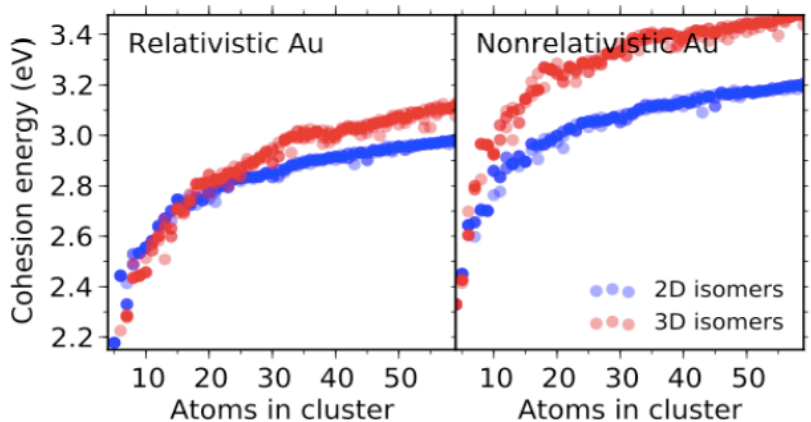

**Fig. S1**. The planar bonding trends in small gold clusters with DFTB. Relativistic (left) and nonrelativistic (right) calculations of the cohesive energy per atom for 2D and 3D low-energy cluster isomers from global optimization (few isomers shown for each cluster size). The transition from 2D to 3D ground state shifts from $N \approx 6$ to $N \approx 15$ upon switching on the relativity; relativity thus favors 2D bonding.

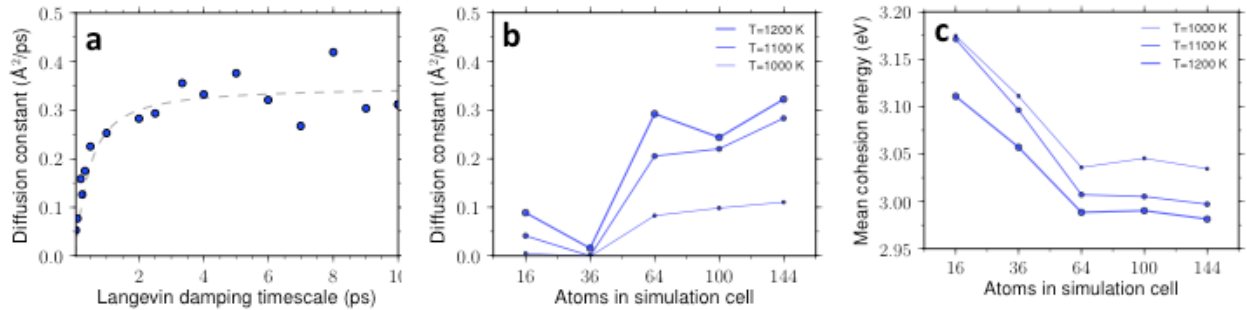

**Fig. S2.** Convergence of simulation parameters. (**A**) Diffusion constant as a function of Langevin damping time for $Au_{64}$ at 1300 K and 4 % area strain. Dashed line is a sketch for a general trend. (**B**) Diffusion constant as a function of simulation cell size at three temperatures. (**C**) Mean cohesive energy per atom as a function of cell size at three temperatures.

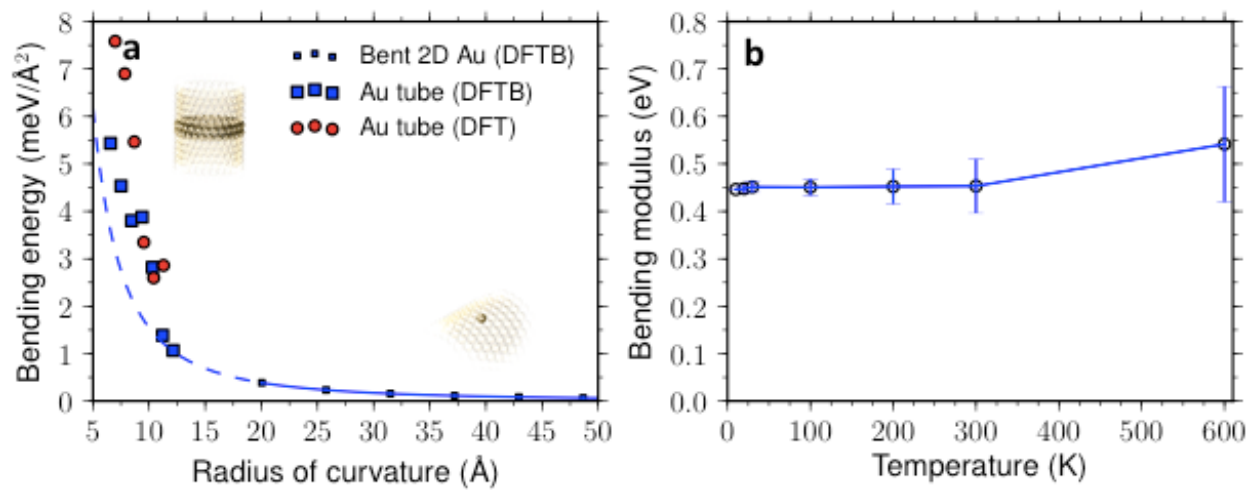

**Fig. S3**. Bending rigidity of 2D Au. (**A**) Bending energy as a function of radius of curvature for bent 2D Au (lower inset; one atom in unit cell using RPBC approach) and Au tubes (upper inset; 28 − 52 atoms in unit cell). A fit above 20 Å yields bending modulus of 0.45 eV (solid line). Extrapolation below 20 Å (dashed line) gives a rough agreement with small-diameter tubes. (**B**) Bending modulus of 2D Au as a function of temperature ($Au_{64}$ RPBC simulations with DFTB). Errorbars are from the error analysis of fitting.

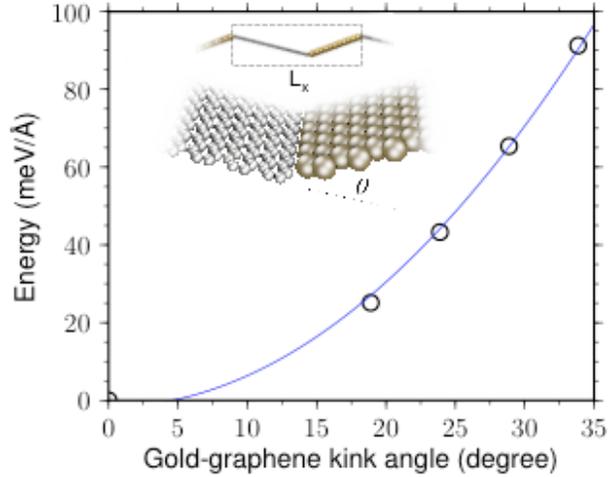

**Fig. S4**. Bending rigidity of Au-C interface. Plot shows the energy density of one Au-C interface as a function of the kink angle $\theta$, as calculated by decreasing the length $L_x$ of a periodic cell that contains two rigid bodies made of Au and C slabs (C: length 3.2 nm; Au: length 2.1 nm; width 2.5 Å; 1×6 k-point sampling). By normalizing the energy to the surface area of the interface, one obtains bending modulus of $\kappa_{Au-C} = 0.54$ eV (solid line), essentially the same as that of 2D Au (Fig. S3).

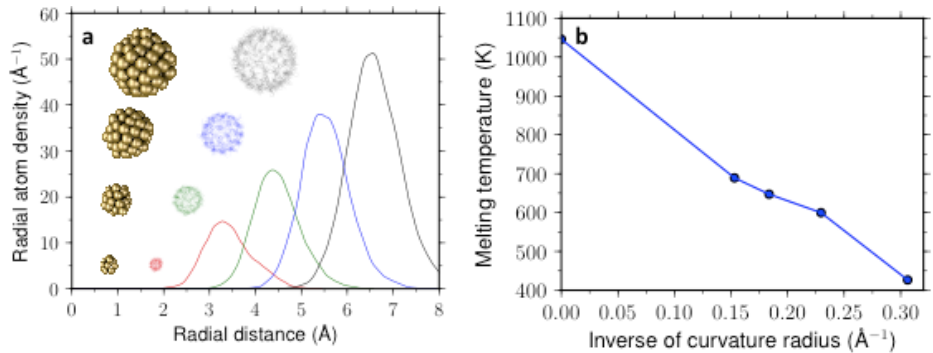

**Fig. S5**. Effect of curvature on melting temperature. (**A**) Radial atom densities of gold cages averaged over 50 ps simulations showing the cavities in the centers of mass. Clusters have 18, 32, 50, and 72 atoms. Insets: snapshots and atom trajectories. (**B**) Melting temperature as a function of inverse of curvature radius. The criterion for liquid was $D > 0.1$ Å$^2$/ps; the melting temperature at zero curvature was adopted from 2D Au simulations.